\documentstyle[12pt]{article}

\begin{document}

\title{ON THE INTERPRETATION OF CYLINDRICALLY SYMMETRIC LEVI-CIVITA
SPACETIME FOR
$0\le\sigma<\infty$}
\author{L. Herrera$^1$\thanks{Postal
address: Apartado 80793, Caracas 1080A, Venezuela.} \thanks{e-mail:
laherrera@telcel.net.ve},
N. O. Santos$^{2,3}$\thanks{e-mail: nos@cbpf.br},
A. F. F. Teixeira$^3$\thanks{e-mail: teixeira@cbpf.br} and
A. Z. Wang$^4$\thanks{e-mail: wang@dft.if.uerj.br}
\\ \\
{\small $^1$Area de F\'{\i}sica Te\'orica, Facultad de Ciencias,}\\
{\small Universidad de Salamanca, 37008 Salamanca, Spain.} \\
{\small $^2$Laborat\'orio Nacional de Computa\c{c}\~ao Cient\'{\i}fica,}\\
{\small 25651-070 Petr\'opolis-RJ, Brazil.}\\
{\small $^3$ LAFEX, Centro Brasileiro de Pesquisas F\'{\i}sicas, 22290-180
Rio de Janeiro-RJ, Brazil.}\\
{\small $^4$Departamento de F\'{\i}sica Te\'orica, Universidade do Estado
do Rio de Janeiro,}\\
{\small 20550-013 Rio de Janeiro-RJ, Brazil.}}
\maketitle

\begin{abstract}
We study Levi-Civita metric for values of its $\sigma$ parameter in the
range $0\le\sigma<\infty$. We show that the value
 $\sigma=1/2$ makes the axial and angular coordinates switch meaning. We
present its geodesics and
a physical source satisfying the energy conditions for all the range of
$\sigma$. This source allows us to obtain  an energy per
unit length which agrees with the behaviour of the geodesics and the fact
that the solution has no event horizon.
\end{abstract}
\pagebreak

\section{Introduction}
The cylindrically symmetric static vacuum solution of Einstein's field
equations was obtained by Levi-Civita (LC) \cite{Levi} in 1919. Ever since
much was written by researchers trying to grasp its
 physical and geometrical interpretations. But this endeavour proved to be
difficult and uncertain. Only in 1958 Marder
\cite{Marder} established that the solution contains two arbitrary
independent parameters usually called $\sigma$ and $a$.
 Understanding the origin, geometry and physics that lies behind these two
parameters is the big challenge to understand the
 solution.
For small values of $\sigma$, as noticed by LC himself, the
corresponding Newtonian field is the external gravitational field produced
by
an infinitely long homogeneous line mass, with $\sigma$
representing the mass per unit length.
In this approximation the parameter $a$ is also associated to the
constant arbitrary potential that exists in the Newtonian solution.
In 1979 Bonnor \cite{Bonnor} pointed out that $a$
is also dressed with a relevant global topological meaning, and cannot be
removed by scale transformations.
There is a series of obstacles and apparently contradictory properties
of $\sigma$ to allow possible interpretations (see a discussion
in \cite{Herrera}). In this article we present some results concerning
$\sigma$ that suggest certain interpretations but, we are aware, collide
with other results. In section 2 we study the nature of the coordinates
LC solution related to the range of different values of $\sigma$ between
0 and $\infty$. Circular geodesics are studied in this same range for
$\sigma$ in section 3. A cylindrical shell source of anisotropic fluid
is matched to LC solution with $0\le\sigma<\infty$ satisfying the energy
conditions in section 4. The definition of energy per unit length is
studied in section 5 for the shell source and compared to the properties
of the geodesics studied in section 3. We end the article with a short
conclusion.

\section{The Levi-Civita metric}

The general static cylindrically symmetric vacuum spacetime satisfying
Einstein's field equations is given by Levi-Civita (LC) metric
\cite{Levi}, which we write in the form
\begin{equation} \label{a1}
ds^2=\varrho^{4\sigma}dt^2-\varrho^{4\sigma(2\sigma-1)}
\left(d\varrho^2+\frac{1}{a^2_m}dm^2\right)
-\frac{1}{a^2_n}\varrho^{2(1-2\sigma)}dn^2,
\end{equation}
where $-\infty<t<\infty$ is the time and $0\le\varrho<\infty$ the
radial coordinates, and $\sigma$, $a_m$ and $a_n$ are constants.
This spacetime clearly has the Killing vectors
$\xi^{\mu}_{(t)}=\delta^{\mu}_t$, $\xi^{\mu}_{(m)}=\delta^{\mu}_m$ and
 $\xi^{\mu}_{(n)}=\delta^{\mu}_n$.
The nature of the coordinates $m$ and $n$, so far unspecified, depends
upon the behaviour of the metric coefficients. Either $a_m$ or $a_n$ can be
transformed away by a scale transformation depending
upon the behaviour of the coordinates $m$ and $n$, thus leaving the metric
with only two independent parameters.
In order to find that behaviour
in a comprehensive way, we first transform the radius $\varrho$ into a
proper
radius $r$ by defining
\begin{eqnarray} \label{a2}
\varrho^{2\sigma(2\sigma-1)}d\varrho=dr,
\end{eqnarray}
so obtaining
\begin{eqnarray} \label{a3}
\varrho=R^{1/\Sigma}, \;\; R=\Sigma r, \;\;
\Sigma=4\sigma^2-2\sigma+1;
\end{eqnarray}
the metric (\ref{a1}) then becomes
\begin{eqnarray} \label{a4}
ds^2=f(r)dt^2-dr^2-h(r)dm^2-l(r)dn^2,
\end{eqnarray}
with
\begin{eqnarray} \label{a5}
f(r)=R^{4\sigma/\Sigma}, \;\;
h(r)=\frac{1}{a^2_m}R^{4\sigma(2\sigma-1)/\Sigma}, \;\;
l(r)=\frac{1}{a^2_n}R^{2(1-2\sigma)/\Sigma}.
\end{eqnarray}

When $\sigma=0$ we have $\Sigma=1$ and considering $a_m=a_n=1$, then
(\ref{a4}) becomes
\begin{eqnarray} \label{a6}
ds^2=dt^2-dr^2-dm^2-r^2dn^2,
\end{eqnarray}
corresponding to a flat four-geometry;
by inspection we interprete $m$ as an axial $z$ coordinate,
and $n$ as an azimuthal $\phi$ coordinate, and write
\begin{eqnarray} \label{a7}
ds^2=dt^2-dr^2-dz^2-r^2d\phi^2.
\end{eqnarray}
It is customary to assume that $\phi$ ranges from 0 to $2\pi$, and
topologically identify these two extremes;
in so doing, the picture of a cartesian two-dimensional plane naturally
follows, in surfaces where $t$ and $z$ are constants.
Equivalently, one may  assume that $\phi$
ranges from $-\infty$ to $\infty$, together with the topological
identification
of every $\phi$ with $\phi+2\pi$.
However, sometimes a two-dimensional flat conical structure with $t$ and $z$
constants appears more suitable to represent the physical situation,
and some modification of the previous assumptions is demanded.
One could simply maintain (\ref{a7}) as the line element,
with the fact that $\phi$ now ranges from 0 to
$2\pi\sin\alpha$, where $\alpha$ is the half-angle of the cone,
together with the topological identification of 0 and $2\pi\sin\alpha$.
However, the most common strategy in these cases is to set
$g_{\phi\phi}=-r^2\sin^2\!\alpha\,d\phi^2$,
while maintaining $\phi$ going from 0 to $2\pi$
with the topological identification of 0 and $2\pi$ as before.
In this paper we often deal with two-dimensional surfaces $t$ and $z$
constants, endowed with rotational symmetry $\xi^{\mu}_{(\phi)}$,
but with radially varying gaussian curvature.
In all circumstances we shall follow the convention that
$-\infty<\phi<\infty$ with the equivalence $\phi\sim\phi+2\pi$.

Concerning the variable $z$ in (\ref{a7}), the most common assumption
is that it goes from $-\infty$ to $\infty$ and that points with
different $z$ are always different;
however, this last statement is an unnecessary topological limitation,
since the topological identification of any $z$ with $z+\zeta$ is admissible
without destroying the flatness of the four-space.
Assuming $z\sim z+\zeta$ promotes compactification of the space in the
$z$ direction (for $\zeta$ finite and nonzero), an occasionally desirable
operation.

Before proceeding, a few words seem worthwhile concerning the
cylindrical coordinates $(r, z, \phi)$ in a curved three-space with
cylindrical
symmetry.
We consider the line element
\begin{eqnarray} \label{a8}
dl^2=dr^2+h(r)dm^2+l(r)dn^2,
\end{eqnarray}
which clearly has the commuting vector fields $\xi^{\mu}_{(m)}$ and
$\xi^{\mu}_{(n)}$;
certainly $r$ is the radial coordinate, but which of the coordinates $m$
and $n$ is the angular $\phi$ and which is the axial $z$?
By analogy with the flat case (where $g_{\phi\phi}=r^2$ and
$g_{zz}=1$),
it seems appropriate to call angular that coordinate whose
metric coefficient vanishes at $r=0$, and call axial the other
coordinate when the corresponding metric coefficient does not vanish at
$r=0$.

We now return to the line element given by (\ref{a4}) and (\ref{a5}),
and consider $0<\sigma<1/2$.
For this range of $\sigma$ we always have $h(r)$ diverging when $r\to 0$,
and always $l(0)=0$; we then visualize $m$ as the axial coordinate $z$,
and $n$ as the angular coordinate $\phi$:
\begin{eqnarray} \label{a9}
ds^2=R^{4\sigma/\Sigma}dt^2-dr^2-R^{-4\sigma(1-2\sigma)/\Sigma}dz^2-
\frac{1}{a^2}R^{2(1-2\sigma)/\Sigma}d\phi^2,
\end{eqnarray}
with possible topological identifications in $z$ and $\phi$.

When $\sigma=1/2$ the two metric coefficients $h$ and $l$ in (\ref{a5})
are constant, unitary.
Then neither $m$ nor $n$ is entitled to be an angular coordinate,
and the three coordinates $(r, m, n)$ are better visualized as cartesian
coordinates $(x, y, z)$. We have, {\it e.g.}, the Rindler flat spacetime
\cite{Silva1}, whose $t$=const sections have planar symmetry:
\begin{eqnarray} \label{a10}
ds^2=z^2dt^2-dx^2-dy^2-dz^2,
\end{eqnarray}
with possible topological identifications in the coordinates $x$ and $y$.

We next consider $1/2<\sigma<\infty$;
in this range of $\sigma$ we always have $h(0)=0$ and $l(r)$ diverging
when $r\to 0$; so we now interprete $m$ as an angular coordinate $\phi$,
and $n$ as an axial coordinate $z$:
\begin{eqnarray} \label{a11} 
ds^2=R^{4\sigma/\Sigma}dt^2-dr^2-R^{-2(2\sigma-1)/\Sigma}dz^2-
\frac{1}{a^2}R^{4\sigma(2\sigma-1)/\Sigma}d\phi^2,
\end{eqnarray}
with possible topological identifications in $z$ and $\phi$, and where we
replaced $a_m$ for $a$. We thus see that the constant $\sigma$ cannot be
chosen arbitrarily, since its value is intimately connected with the
boundary conditions on the $z$ axis and on the radially far regions.

The Kretschmann scalar
${\cal R} =R_{\alpha\beta\gamma\delta}R^{\alpha\beta\gamma\delta}$
for the metric (\ref{a4}) is \cite{Wang}
\begin{equation} \label{a12}
{\cal R}=\frac{64\sigma^2(2\sigma-1)^2}{\Sigma^3r^4};
\end{equation}
from (\ref{a12}) we see that the spacetime (\ref{a4}) is locally flat
only for $\sigma=0,1/2$ and $\infty$.
In section {\bf 3} we match (\ref{a9}) and (\ref{a11}) to a cylindrical
anisotropic shell of matter.
The interior of the shell cylinder is assumed to be Minkowski
spacetime, hence there the coordinates have a well defined meaning.
The matching condition, of the continuity of the metrics, relates the
respective
metric coefficients for each coordinate, in particular $g_{zz}$ and
$g_{\phi\phi}$, thus giving a further support to the coordinates chosen
in (\ref{a9}) and (\ref{a11}).

\section{Circular
geodesics}
For the circular geodesics \cite{Silva} we have $\dot{r}=\dot{z}=0$
and $g_{\phi\phi,r}\dot{\phi}^2+g_{tt,r}\dot{t}^2=0$ where the dot
stands for differentiation with respect to $s$. The angular velocity
$\omega$ of a particle moving along a geodesics is
$\omega=\dot{\phi}/\dot{t}$ and its tangential velocity $W^{\mu}$ is
given by $W^{\phi}=\omega/\sqrt{g_{tt}}$ which is the only non null
component.

In the case $0\le\sigma<1/2$, from (\ref{a9}) we obtain
\begin{eqnarray} \label{a13}
\omega^2=\frac{2\sigma}{1-2\sigma}a^2R^{2(4\sigma-1)/\Sigma},
\end{eqnarray}
\begin{eqnarray} \label{a14}
W^2=\frac{2\sigma}{1-2\sigma};
\end{eqnarray}
and, in the case $1/2<\sigma<\infty$, from (\ref{a11}) we have
\begin{eqnarray} \label{a15}
\omega^2=\frac{1}{2\sigma-1}a^2R^{8\sigma(1-\sigma)/\Sigma},
\end{eqnarray}
\begin{eqnarray} \label{a16}
W^2=\frac{1}{2\sigma-1}.
\end{eqnarray}
It is worthwhile noting that for a given system (i.e. a fixed $\sigma$)
the squared velocity $W^2$ is the same for all circular geodesics,
in agreement with the corresponding Newtonian gravitation.

We see from (\ref{a14}) that $W$ monotonically increases with $\sigma$,
that is from $\sigma=0$ producing $W=0$, to $\sigma=1/4$ attaining
$W=1$ (the speed of light),
and finally $\sigma=1/2$ producing geodesics with $W=\infty$.
With $\sigma$ increasing beyond 1/2, we note from (\ref{a16}) that $W$
diminishes,
attaining $W=1$ for $\sigma=1$ and $W=0$ for $\sigma=\infty$.
In other words, the circular geodesics are timelike when either
$0<\sigma<1/4$ or $\sigma>1$, are lightlike when $\sigma=1/4$ or $\sigma=1$,
and are spacelike when $1/4<\sigma<1$.

These facts suggest that, while $\sigma$ increases from zero to $1/2$,
the effective energy density per unit length, $\mu$, of the line source
that produces the spacetime (\ref{a9}) increases too.
So that for a test particle to remain in a circular motion its velocity
has to increase in order to make a balance between the attracting
gravitational
force, that increases with $\mu$, and the centrifugal force,
that increases with $W$.
On the other hand, when $\sigma$ further increases from 1/2 to $\infty$
in the spacetime (\ref{a11}), it appears that the effective energy
per unit length monotonically decreases to zero value.

\section{Matching LC spacetime to a cylindrical shell
source}
Here we follow the same procedure as in \cite{Wang}.
We consider an infinitely thin cy\-lin\-drical shell of anisotropic
fluid matter with a finite radius and we match it to the exterior LC
spacetime given by
either (\ref{a9}) or (\ref{a11}). For simplicity we shall  assume that
the source is static.
For the interior of the shell cylinder we assume Minkowski spacetime,
since it is the only static spacetime deprived of energy density
\cite{Apostolatos}.

In order to do the matching we require the continuity of the metric
coefficients across the shell \cite{Taub},
allowing us to obtain the most general anisotropic shell fluid source.
Using the same coordinate system as in (\ref{a9}) or (\ref{a11}), we
have for the interior $0\le r< r_0$ of the shell cylinder with radius
$r=r_0$
the Minkowski spacetime,
\begin{equation} \label{a17}
ds^2_-=dt^2-dr^2-dz^2-r^2d\phi^2.
\end{equation}
Indices $-$ and $+$ refer to interior and exterior spacetimes, respectively.
In order to have the general matching at $r=r_0$ satisfied for the
exterior LC metric, we make a reparametrization of $t$ and $z$
like\begin{equation} \label{a18}
t\to\frac{t}{A}, \;\; z\rightarrow\frac{z}{B},
\end{equation}
where $A$ and $B$ are constants. Then (\ref{a9}) with (\ref{a18})
becomes, for
$0\le\sigma\le 1/2$,
\begin{equation} \label{a19}
ds^2_+=\frac{1}{A^2}R^{4\sigma/\Sigma}dt^2-dr^2-\frac{1}{B^2}
R^{4\sigma(2\sigma-1)/\Sigma}dz^2
-\frac{1}{a^2}R^{2(1-2\sigma)/\Sigma}d\phi^2,
\end{equation}
and (\ref{a11}) with (\ref{a18}), for $1/2\le\sigma<\infty$,\begin{equation} \label{a20}
ds^2_+=\frac{1}{A^2}R^{4\sigma/\Sigma}dt^2-dr^2-\frac{1}{B^2}R^{2(1-2\sigma)
/\Sigma}dz^2-\frac{1}{a^2}R^{4\sigma(2\sigma-1)/\Sigma}d\phi^2.
\end{equation}
Then, considering the junction condition \cite{Taub}
\begin{equation} \label{a21}
g^+_{\mu\nu}\mid_{r_0}=g^-_{\mu\nu}\mid_{r_0},
\end{equation}
we obtain from (\ref{a17}) and (\ref{a19}), for $0\le\sigma\le 1/2$,
\begin{equation} \label{a22}
A=R^{2\sigma/\Sigma}_0, \;\; B=R^{2\sigma(2\sigma-1)/\Sigma}_0, \;\;
ar_0=R^{(1-2\sigma)/\Sigma}_0,
\end{equation}
where $R_0=\Sigma r_0$; and from (\ref{a17}) and (\ref{a20}), for
$1/2\le\sigma<\infty$,
\begin{equation} \label{a23}
A=R^{2\sigma/\Sigma}_0, \;\; B=R^{(1-2\sigma)/\Sigma}_0, \;\;
ar_0=R^{2\sigma(2\sigma-1)/\Sigma}_0.
\end{equation}
Taub has shown \cite{Taub} that if (\ref{a21}) is satisfied then the first
derivatives of the metric are { in general} discontinuous across $r=r_0$,
giving rise to a shell of matter. Following him,
\begin{equation} \label{a24}
g^+_{\mu\nu,\lambda}\mid_{r_0}-g^-_{\mu\nu,\lambda}\mid_{r_0}=
n_{\lambda}b_{\mu\nu},
\end{equation}
where $n_{\lambda}$ is the normal to the hypersurface $r=r_0$, directed
outwards, giving
$n_{\lambda}=\delta^r_{\lambda}$.
{}From (\ref{a24}) we calculate $b_{\mu\nu}$ and obtain the energy momentum
tensor $T_{\mu\nu}$ of the shell, which is given by
\begin{equation} \label{a25}
T_{\mu\nu}=\tau_{\mu\nu}\delta(r-r_0),
\end{equation}
where
\begin{equation} \label{a26}
\tau_{\mu\nu}=\frac{1}{16\pi}[b(ng_{\mu\nu}-n_{\mu}n_{\nu})
+n_{\lambda}(n_{\mu}b^{\lambda}_{\nu}+n_{\nu}b^{\lambda}_{\mu})-nb_{\mu\nu}-
n_{\lambda}n_{\delta}b^{\lambda\delta}g_{\mu\nu}],
\end{equation}
and where $\delta(r-r_0)$ denotes the Dirac delta function,
$n=n_{\lambda}n^{\lambda}$, and $b=b^{\lambda}_{\lambda}$.
Considering (\ref{a17}) and (\ref{a19}) we obtain from (\ref{a24})
the nonvanishing components of $b_{\mu\nu}$ for $0\le\sigma\le 1/2$,
\begin{equation} \label{a27}
b_{\mu\nu}=\frac{4\sigma}{R_0}diag(1, 0, 1-2\sigma, 2\sigma r_0^2),
\end{equation}
where the order of the coordinates is $(t, r, z, \phi)$;
and from (\ref{a17}), (\ref{a20}), and (\ref{a24}) we find for
$1/2\le\sigma<\infty$
\begin{equation} \label{a28}
b_{\mu\nu}=\frac{2}{R_0}diag(2\sigma, 0, 2\sigma-1, r_0^2).
\end{equation}
With (\ref{a27}) and (\ref{a28}) substituting into (\ref{a26}) we can
write the shell energy momentum ancillary tensor as
\begin{equation} \label{a29}
\tau_{\mu\nu}=diag(\rho, 0, p_z, r_0^2p_\phi),
\end{equation}
where $\rho$ is the energy density and
$p_z$ and $p_{\phi}$ are the pressures in the $z$ and $\phi$ directions
respectively.

For $0\le\sigma\le 1/2$ these quantities measure
\begin{equation} \label{a30}
\rho=\frac{\sigma}{4\pi R_0}, \;\; p_z=\frac{\sigma(1-2\sigma)}{4\pi R_0},
\;\; p_{\phi}=\frac{\sigma^2}{2\pi R_0},
\end{equation}
while for $1/2\le\sigma<\infty$ they are
\begin{equation} \label{a31}
\rho=\frac{\sigma}{4\pi R_0}, \;\; p_z=\frac{2\sigma-1}{8\pi R_0}, \;\;
p_{\phi}=\frac{1}{8\pi R_0}.
\end{equation}
It can easily be shown that the anisotropic fluids described by (\ref{a30})
and (\ref{a31}) satisfy the weak, strong and dominant energy conditions
\cite{Hawking}. In particular, neither the linear mass density $\rho$ nor
the anisotropic pressures $p_z$ and $p_{\phi}$ are negative in any of the
equations (30) and (31).

\section{The energy per unit length of the shell}

In the Newtonian limit of the LC spacetime $\sigma$ can be interpreted as
the energy per unit length of the source. Considering Israel's definition of
energy density per unit length \cite{Israel} we obtain, for both
(\ref{a30}) and (\ref{a31}),
\begin{equation} \label{a32}
\mu=\int^\infty_0\int^{2\pi}_0(\rho+p_z+p_{\phi})\delta(r-r_0)\sqrt{-g}
drd\phi= \frac{\sigma}{\Sigma},
\end{equation}
where $g$ is the determinant of the metric.
{}From (\ref{a32}) we see that for small $\sigma$ we have
$\mu\approx\sigma$, which is consistent with the Newtonian limit.
As $\sigma$ increases $\mu$ increases too, reaching a maximum at
$\mu_{max}=\sigma=1/2$; then $\mu$ starts diminishing with further
increase of $\sigma$, becoming $\mu=0$ for $\sigma\rightarrow\infty$.

These results are consistent with (\ref{a14}) and (\ref{a16}),
showing that the tangential speed steadily increases with $\sigma$ and
$\mu$ up to a maximum $W\rightarrow\infty$ at $\sigma=\mu_{max}=1/2$, and
then decreases with increasing $\sigma$ and decreasing $\mu$ up to $W=0$,
when also $\mu=0$.
Furthermore, when $W=1$ the circular geodesics are null. From
(\ref{a14}) and (\ref{a16}) we can see that this corresponds to
$\sigma=1/4$ and $\sigma=1$, for which we have $\mu(1/4)=\mu(1)=1/3$. These
properties between $\mu$ and $W$ can be seen too from the expression of
$\mu$ in terms of $W$ which is, for $\sigma<1/2$, from (\ref{a14}), as well
as for $\sigma>1/2$, from (\ref{a16}),
\begin{equation}
\label{a34}
\mu=\frac{W^2(1+W^2)}{2(1+W^2+W^4)}.
\end{equation}

It is worth noticing that the relations above between $\omega$, $\sigma$
and $\mu$ are consistent with the behaviour
of a gyrosocope moving along a circular path (not necessarily a geodesic)
with angular velocity $\omega$.

Indeed, using the Rindler-Perlick method \cite{Rindler}, it is not
difficult to find the rate of precession $\Omega$ of such gyroscope for
the line element
(\ref{a4}),
\begin{equation} \label{a34}
\Omega=\frac{\omega(l^{\prime}f-lf^{\prime})}{2\sqrt{fl}(f-\omega^2l)},
\end{equation}
where $l=g_{\phi\phi}$ and $\omega$ is any angular velocity of the gyroscope
around the line
source.
For $0\le\sigma\le 1/2$ we have from (\ref{a9}),
\begin{equation} \label{a35}
\Omega=\frac{a\omega(1-4\sigma)R^{-2\sigma(1+2\sigma)/\Sigma}}{a^2-\omega^2R
^{2(1-4\sigma)/\Sigma}};
\end{equation}
whereas for $1/2\le\sigma<\infty$ we have from (\ref{a11}),
\begin{equation} \label{a36}
\Omega=\frac{4a\omega
\sigma(\sigma-1)R^{-(1+2\sigma)/\Sigma}}{a^2-\omega^2R^{8\sigma(\sigma-1)/\S
igma}}.
\end{equation}
Thus we see that the gyroscope is locked at the lattice, $\Omega=0$, for
$\sigma=1/4$ and $1$ respectively, as expected for null paths
\cite{Rindler}, and both values of
$\sigma$ produce from (\ref{a32}) $\mu(1/4)=\mu(1)=1/3$.

Also, for $\sigma=0$ we obtain the Thomas precession (modified by the
effect of $a$)
\begin{equation} \label{a37}
\Omega=\frac{a\omega}{a^2-\omega^2r^2};
\end{equation}
and similarly for $\sigma\gg 1$,
\begin{equation} \label{38}
\Omega\approx\frac{a\tilde{\omega}}{a^2-\tilde{\omega}^2r^2},
\end{equation}
where $\tilde{\omega}=4\omega\sigma^2$.
However, for $\sigma=1/2$ the situation is not so clear (see \cite{Herrera}
and the discussion below).

Also observe that $\mu$ has only one maximum at $\sigma=1/2$ , and the
maximum is finite, with the value $=1/2$.
Then, since in cylindrical sources no black holes are formed, one might
conclude that the minimum mass per unit length
to form a black hole satisfies the constraint $\mu>1/2$.
Indeed, according to our present understanding of the black hole physics,
there should exist a lower mass limit to form them. For example, it is
accepted that the mass of a  neutron star is
between $1.2 M_{\odot}$ and
$1.7M_{\odot}$, where $M_{\odot}$ denotes the solar mass \cite{Gled97}.
When the mass of a star is $M \gg 1.7M_{\odot}$, it might exist
only in the state of a black hole.

There are two other expressions for mass per unit length besides Israel's
(\ref{a32}). One is given by Marder \cite{Marder}, which is
\begin{eqnarray}
\label{a35}
\mu_M=\int^{\infty}_{0}\int^{2\pi}_{0}\rho\delta(r-r_0)\sqrt{g_{(2)}}dr
d\phi,
\end{eqnarray}
where $g_{(2)}$ is the determinant of the induced metric on the
2-surface defined by $t=z=$const. The other definition is given by
Vishveshwara and Winicour \cite{Vishveshwara}, based on the Killing vectors
of
time translation $\xi^{\mu}_{(0)}=\delta^{\mu}_t$ and rotation
$\xi^{\mu}_{(3)}=\delta^{\mu}_{\phi}$, which is
\begin{eqnarray}
\mu_{VW}=-\frac{1}{2\tau}(\lambda_{33}\lambda_{00,\tau}-\lambda_{03}\lambda_
{03,\tau}),
\end{eqnarray}
where
\begin{eqnarray}
\lambda_{00}=\xi^{\mu}_{(0)}\xi_{\mu(0)}, \;\;
\lambda_{03}=\xi^{\mu}_{(0)}\xi_{\mu(3)}, \;\;
\lambda_{33}=\xi^{\mu}_{(3)}\xi_{\mu(3)},
\end{eqnarray}
\begin{equation}
\tau^2=-2(\lambda_{00}\lambda_{33}-\lambda^2_{03}).
\end{equation}
Using (\ref{a19},\ref{a30}) or (\ref{a20}),\ref{a31}) we obtain
\begin{equation}
\mu_M=\frac{\sigma}{2\Sigma}, \;\; \mu_{VW}=\sigma;
\end{equation}
we see that $\mu_M$ does not produce the Newtonian limit, and $\mu_{VW}$
does not explain the circular geodesics behaviour. Hence we discard both
definitions. (In \cite{Wang} the expressions (15) and (16) should be
interchanged with their respective analises.)

\section{Conclusion}
We have presented the cylindrically symmetric static vacuum solution of
Einstein's field equations, obtained by LC in its general
form, by only specifying the time and radial coordinates and not specifying
the other two.
We showed that the nature of the two coordinates is closely linked with the
range of $\sigma$.
There are two ranges, $0\le\sigma<1/2$ and $1/2<\sigma<\infty$, where the
coordinates,
being $z$ and $\phi$, switch their nature \cite{Yasuda00}. We calculated the
circular geodesics for these
two ranges and it appears, from their behaviour, that the energy per unit
length increases by
increasing $\sigma$ up to $1/2$ and then diminishes while $\sigma$
increases to $\infty$.

We matched a shell source to LC solution satisfying the energy conditions
for the whole range of $\sigma$.
The energy per unit length $\mu$ calculated from the definition given by
Israel reproduces the behaviour of
the circular geodesics and, furthermore, while producing a maximum for
$\mu$ suggests a possible explanation for
the non existence of event horizons in LC spacetime. Of course, we are aware
that
$\mu$ is model dependent and cannot
be given a general meaning but none the less it satisfies part of the
expected properties.

Some questions, however, remain unanswered, which leaves the puzzle
incomplete.

Indeed, observe that for $\sigma=1/2$ the energy density $\rho$ as well as
$p_{\phi}$ and $\mu$ are nonvanishing, but the
spacetime is flat. Now, if the source would be a plane (as seems to be the
case in \cite{BS}) then one could invoke
the equivalence principle to explain the vanishing of the Riemann tensor.
However in our case the proper radius of the cylinder
(unlike the case analyzed in \cite{BS}) remains constant and finite for any
value of $\sigma$. So the question is: why does a cylinder
with positive energy density and pressure distribution produce vanishing
curvature? This last question, together with the
fact that a gyroscope moving along the $\phi$-coordinate in the LC
spacetime with $\sigma=1/2$ behaves in an unexpected way (see
\cite{Herrera}), brings out the difficulties in interpreting the
$\sigma=1/2$
case as due to a cylinder, in spite of the fact that
the source presented in section 4 is physically satisfactory.

The situation described above might have its origins in the following two
facts:
\begin{enumerate}
\item The source is a shell and therefore the second fundamental form is
discontinuous across the boundary surface. It is then
unclear whether  demanding both fundamental forms to be continuous is
compatible with a reasonable cylindrical source
or if, instead, it leads
always to the same results as in \cite{BS} when $\sigma=1/2$. Indeed, the
use of a shell source, although very useful for its simplicity,
may screen in some cases, relevant aspects of the problem. On the other
hand the discussion on the basis  of a general (not related to a specific
equation of state)
source satisfying both Darmois conditions, is presently out of our reach.
\item Axial and angular coordinates switch meaning at $\sigma=1/2$, so it
is reasonable to ask if for  this value of $\sigma$, $r=r_{0}$
describes a cylinder. In fact, as commented above equation (10), in this
case the three spatial coordinates are better visualized as cartesian
coordinates.
Therefore one is leaned to identify the topology of the source with that of
a plane rather than a cylinder.
\end{enumerate}
At any rate it is clear that this point requires further discussion.

\section{Acknowledgment}
NOS is grateful to Professor Alberto Santoro for his kind hospitality at
the Centro Brasileiro de Pesquisas F\'{\i}sicas, where this work
was developed.The financial assistance from
CNPq (NOS, AZW) and FAPERJ (AZW) is also gratefully acknowledged.

\end{document}